
\magnification=\magstep1
\hoffset=0.1truecm
\voffset=0.1truecm
\vsize=23.0truecm
\hsize=16.25truecm
\parskip=0.2truecm
\def\newpage{\vfill\eject}

\def\ntot{N}

\def\jfun{J}
\def\kfun{K}
\def\lfun{L}

\def\fh1{ \bigl[- {d h \over dx} \bigr] }
\def\h2{ \bigl[- {1 \over h} {d h \over dx} \bigr] }
\def\f1{\Bigl[ \int_0^1 f dx \Bigr]}

\def\mp{ {M_{\rm pl}}}
\def\lft{ {\lambda_{FT} } }
\def\rat{ {\cal R} }
\def\rave{ {\langle {\cal R} \rangle } }

\def\gwper{ {\delta_T} }
\def\index{ {\alpha} }
\def\vr{ {v} }
%
%
\centerline{\bf THE SCALAR FIELD POTENTIAL IN INFLATIONARY MODELS: }
\smallskip
\centerline{\bf RECONSTRUCTION AND FURTHER CONSTRAINTS}
\bigskip
\centerline{\bf Fred C. Adams and Katherine Freese}
\bigskip
\centerline{\it Physics Department, University of Michigan}
\centerline{\it Ann Arbor, MI 48109, USA}
\bigskip
\centerline{\it submitted to Physical Review {\rm D}: 6 January, 1994}

\vskip 0.40truein
\goodbreak
\centerline{\bf Abstract}
\medskip

In this paper, we present quantitative constraints on the scalar
field potential for a general class of inflationary models.
(1) We first consider the reconstruction of the inflationary
potential for given primordial density fluctuation spectra.
Our work differs from previous work on reconstruction in that
we find a semi-analytic solution for the potential for the case of
density fluctuations with power-law spectra.  In addition, for
the case of more general spectra, we show how constraints on the
density fluctuation spectra imply corresponding constraints on
the potential.  We present a series of figures which show how the
shape of the potential depends on the shape of the perturbation
spectrum and on the relative contribution of tensor modes.
(2) We show that the average ratio $\rave$ of the amplitude of
tensor perturbations (gravity wave perturbations) to scalar density
perturbations is bounded from above:  $\rave \le$ 1.6.  We also show
that the ratio $\rave$ is proportional to the change $\Delta \phi$
in the field: $\rave \approx 0.42 \Delta \phi/\mp$.  Thus, if
tensor perturbations are important for the formation of structure,
then the width $\Delta \phi$ must be comparable to the Planck mass.
(3) We constrain the change $\Delta V$ of the potential and the
change $\Delta \phi$ of the inflation field during the portion of
inflation when cosmological structure is produced.  We find both
upper and lower bounds for $\Delta \phi$ and for $\Delta V$.
In addition, these constraints are then used to derive a bound on
the scale $\Lambda$, which is the scale of the height of the
potential during the portion of inflation when cosmological
perturbations are produced;  we find $\Lambda \le 10^{-2} \mp$.
Thus, the last $\sim$60 e-foldings of inflation must take place
after the GUT epoch.  This bound on $\Lambda$, although comparable
to those found previously, is found here using different methods.
(4) In an earlier paper, we defined a fine-tuning parameter
$\lft$ $\equiv$ $\Delta V/ (\Delta \phi)^4$ and found an upper
bound for $\lft$.  In this paper, we find a lower bound on
$\lft$.  The fine-tuning parameter is thus constrained to lie
in the range $6 \times 10^{-11}$ $(\Lambda/10^{17} {\rm GeV})^8$
$\le \lft \le$ $10^{-7}$.
(5) Finally, we consider the effects of requiring a
non-scale-invariant spectrum of perturbations (i.e., with
spectral index $n \ne$1) on the fine-tuning parameter $\lft$.
For spectral indices $n$ less than unity, the upper bound on
the fine-tuning parameter $\lft$ becomes more restrictive than
the $n=0$ case by a factor ${\cal F} \sim$ 2 -- 5.

\vskip 0.3truein
PAC Numbers: 98.80.--k, 98.80.Cq, 12.10.Dm

\newpage
\centerline{\bf I. INTRODUCTION}
\medskip

The inflationary universe model [1] provides an elegant means of
solving several cosmological problems, including the horizon problem,
the flatness problem, and the monopole problem.  In addition, quantum
fluctuations produced during the inflationary epoch may provide the
initial conditions required for the formation of structure in the
universe.  During the inflationary epoch, the energy density of the
universe is dominated by a (nearly constant) vacuum energy term
$\rho \simeq \rho_{vac}$, and the scale factor $R$ of the universe
expands superluminally (i.e., ${\ddot R} > 0$).  If the time interval
of accelerated expansion satisfies $\Delta t \ge 60 R/ {\dot R}$, a
small causally connected region of the universe grows sufficiently
to explain the observed homogeneity and isotropy of the universe,
to dilute any overdensity of magnetic monopoles, and to flatten the
spatial hypersurfaces (i.e., $\Omega \rightarrow 1$).  In most models,
the vacuum energy term is provided by the potential of a scalar field.
In this paper, we present constraints on this scalar field potential
for a general class of inflationary models.  This present work extends
the results of a previous paper [2] where we quantified the degree
of fine tuning required for successful inflationary scenarios.

In the original model of inflation [1], now referred to as Old
inflation, the universe supercools to a temperature $T \ll T_C$ during
a first-order phase transition with critical temperature $T_C$.  The
nucleation rate for bubbles of true vacuum must be slow enough that
the Universe remains in the metastable false vacuum long enough for
the required $\sim 60$ e-foldings of the scale factor. Unfortunately,
the Old inflationary scenario has been shown to fail [3] because the
interiors of expanding spherical bubbles of true vacuum cannot
thermalize properly and produce a homogeneous radiation-dominated
universe after the inflationary epoch.

In order to overcome the ``reheating problem'' of the Old inflationary
scenario, a new class of inflationary models was developed [4, 5, 6].
In this class of models, the diagram of the effective potential (or
free energy) of the inflation field $\phi$ has a very flat plateau
and the field evolves sufficiently slowly for inflation to take place
(i.e., the field evolves by ``slowly rolling'' off the plateau). In
these models, the phase transition can be second order or weakly first
order.  Many inflationary models which are currently under study are
of this latter type, e.g., New Inflation [4, 5], Chaotic Inflation
[6], and Natural Inflation [7].  The evolution of the field $\phi$
is determined by the equation of motion
$$\ddot\phi + 3H\dot\phi + \Gamma\dot\phi + {dV \over d\phi} =
     0 \ , \eqno(1.1)$$
where $H$ is the Hubble parameter and $V$ is the potential.
The $\Gamma {\dot \phi}$ term determines the decay rate of the
$\phi$ field at the end of inflation (see, e.g., Ref. [8]). In
this equation of motion, spatial gradient terms have been neglected
(gradients are exponentially suppressed during the inflationary
epoch).

In most studies of inflation, the field $\phi$ is assumed to be
``slowly rolling'' during most of the inflationary epoch.  The
slowly rolling approximation means that the motion of the inflation
field is overdamped, $\ddot \phi = 0$, so that equation [1] becomes
a first order equation; the $\Gamma {\dot \phi}$ term is also generally
negligible during this part of inflation.  Thus, the motion is
controlled entirely by the force term ($dV / d\phi$) and the viscous
damping term ($3H \dot \phi$) due to the expansion of the universe.
Near the end of the inflationary epoch, the field approaches the
minimum of the potential (i.e., the true vacuum) and then oscillates
about it, while the $\Gamma {\dot \phi}$ term gives rise to particle
and entropy production.  In this manner, a ``graceful exit'' to
inflation is achieved.

For completeness, we note that workable models of inflation which use
a first order phase transition have been proposed, notably Extended
Inflation [9] and Double Field Inflation [10].  However, these models
require an additional ``slowly rolling'' field in order to complete
the phase transition.

All known versions of inflation with slowly rolling fields produce
density fluctuations, which tend to be overly large unless the
potential for the slowly rolling field is very flat.  In particular,
these models produce (scalar) density fluctuations [11] with
amplitudes given by
$${\delta \rho \over \rho}\bigg|_{\rm scalar} \simeq {1 \over 10}
\, {H^2 \over \dot\phi} \ \ , \eqno(1.2)$$
where ${(\delta \rho /\rho)}|_{\rm scalar}$ is the amplitude of a
density perturbation when its wavelength crosses back inside the
horizon (more precisely, the Hubble length) after inflation, and the
right hand side is evaluated at the time when the fluctuation crossed
outside the Hubble length during inflation.  The expression [1.2]
applies to any inflationary model which has a slowly rolling field
$\phi$.  The quantum fluctuations in the motion of the field $\phi$
cause the hypersurface of the phase transition to be nonuniform and
result in density perturbations with magnitude given by Eq.~[1.2].
For the case of Extended or Double Field Inflation, these
perturbations are superimposed with the perturbations caused
by the collisions of bubbles.

In addition to the scalar perturbations described above,
inflationary models can also produce tensor perturbations (gravity
wave perturbations -- see Ref. [12]).  The amplitude of these
perturbations is given by
$${\delta \rho \over \rho} \Bigg|_{\rm GW} \, = { \sqrt{32 \pi} \over
30} \, {H \over \mp} \, , \eqno(1.3)$$
where the right hand side is again evaluated at the time when the
fluctuation crossed outside the Hubble length during inflation
(and where $\mp$ is the Planck mass).

The allowable amplitude of these perturbations is highly
constrained by measurements of the isotropy of the microwave
background. On scales of cosmological interest, these measurements
[13] indicate that
$${\delta \rho \over \rho}\bigg|_{\rm hor} \le \delta \approx
     2 \times 10^{-5} \, .  \eqno(1.4)$$
In this expression, the left hand side represents the total amplitude
of perturbations produced by inflation, i.e., both scalar
perturbations (Eq. [1.2]) and tensor perturbations (Eq. [1.3]);
note that the two types of perturbations add in quadrature.
The right hand side of equation [1.4] represents the experimental
measurements (both detections and limits) of the cosmic microwave
background.  In general, these measurements are a function of the
observed size scale (or angular scale). All measurements
(to date) on different size scales are roughly consistent with
an approximate value $\delta \approx$ $2 \times 10^{-5}$;
details of scale dependence will be considered later.

For the general class of inflationary models with slowly rolling
fields, the coupled constraints that the universe must inflate
sufficiently and that the density perturbations must be sufficiently
small require the potential $V (\phi)$ to be very flat [2, 14].
In a previous paper [2], we derived upper bounds on a
``fine-tuning parameter'' $\lft$, defined by
$$\lft \equiv { \Delta V \over (\Delta \phi)^4 } \ \ , \eqno(1.5)$$
where $\Delta V$ is the decrease in the potential $V(\phi)$
during a given portion of the inflationary epoch and
$\Delta \phi$ is the change in the value of the field $\phi$
over the same period.  In this paper, we define $\Delta V$ and
$\Delta \phi$ over the portion of inflation where cosmic
structure is produced; as discussed below, this portion of
inflation corresponds to the $N$ $\approx$ 8 e-foldings
which begin roughly 60 e-foldings before the end of inflation.
The parameter $\lft$ is the ratio of the height of the potential
to its $({\rm width})^4$ for the part of the potential involved
in the specified time period; $\lft$ thus measures the required
degree of flatness of the potential.  In Ref. [2], we found
that $\lft$ is constrained to be very small for all inflationary
models which satisfy the density perturbation constraint and which
exhibit overdamped motion; in particular, we obtained the bound
$$\lft \le \, {2025 \over 8} \, \delta^2 \approx
10^{-7} \, . \eqno(1.6{\rm a})$$
We also showed that if the potential is a quartic polynomial with
the quartic term in the Lagrangian written as ${1 \over 4}$
$\lambda_q \phi^4$, then a bound on $\lft$ implies a
corresponding bound on $\lambda_q$; specifically,
$$|\lambda_q| \le 36 \lft  \ . \eqno(1.6{\rm b})$$
Thus, the bound of equation [1.6a] implies that the quartic
coupling constant must be extremely small.

In this paper we continue a quantitative study of the constraints on
the scalar-field potential for models of inflation that have a slowly
rolling field.  In the first part of this paper, we consider the
reconstruction of the inflationary potential for given primordial
density fluctuation spectra.  This reconstruction process has already
been considered by many recent papers [15].  In this paper, we show
that for the case of density fluctuations with power-law spectra, the
reconstruction of the inflationary potential can be done
semi-analytically and we find the corresponding semi-analytic
solutions (see Eq. [3.12]).  For the more general case, we show how
constraints on the density fluctuation spectra imply corresponding
constraints on the potential.

Our results show how the shape of the potential depends on the
perturbation spectrum and on the relative contribution of tensor
modes and scalar perturbations (see Figures 1 -- 5).  For the case in
which tensor perturbations produce a substantial contribution to the
total (e.g., in Figure 1), the potentials $V(\phi)$ are concave
upward for all of the spectral indices $n=0.5 - 1$ considered here.
For the opposite case in which tensor modes are negligible
(e.g., in Figure 4), the potentials are concave downward and somewhat
like the cosine potential used in models of Natural Inflation [7].
Figure 5 shows a cosine potential which has been fit to the reconstructed
potential for a particular case with little contribution from tensor
modes (see Sec.  III).  Thus, for perturbation spectra with little
contribution from tensor modes (and moderate departures from scale
invariance), the reconstructed potential looks very much like a
cosine potential.

In the next part of this paper, we present further constraints on the
inflationary potential.  In particular, we constrain both $\Delta V$
and $\Delta \phi$ individually.  We show that both upper and lower
bounds exist for $\Delta \phi$ and for $\Delta V$ (see equations
[4.14], [4.15], and [4.33]).  In addition, these
constraints are used to derive a
bound on the scale $\Lambda$, i.e., the scale of the height of the
potential during the portion of inflation when cosmological
perturbations are produced; we obtain the bound $\Lambda \le 10^{-2} \mp$.
Thus, the final $\sim$60 e-foldings of inflation must take place after
the GUT epoch.  Although found by different methods, this bound on
$\Lambda$ is comparable to those found previously [16, 17, 18, 19].

Next, we show that the average ratio $\rave$ of the amplitude of
tensor perturbations (gravity wave perturbations) to scalar density
perturbations is bounded from above: $\rave \le$ 1.6.  Thus, tensor
perturbations cannot be larger than scalar perturbations by an
arbitrarily large factor.  We also show that the ratio $\rave$ is
proportional to the change $\Delta \phi$ in the field; in particular,
we find that $\rave \approx 0.42 \Delta \phi/\mp$.  Thus, if tensor
perturbations are important for the formation of cosmological structure,
then the width $\Delta \phi$ must be comparable to the Planck mass.

Finally, we consider bounds on the fine-tuning parameter $\lft$.
We find a lower bound on $\lft$ (see equation [5.1]).
We also consider the effects of requiring a non-scale-invariant
spectrum of perturbations (i.e., with spectral index $n \ne$1) on the
fine-tuning parameter $\lft$.  We show that for $n < 1$, the bound on
the fine-tuning parameter $\lft$ becomes more restrictive than the
$n=0$ case (which is effectively the case considered in Ref. [2]).

The constraints presented in this paper apply to inflationary
models involving one or more scalar fields that are minimally
coupled to gravity, and which satisfy three conditions.
First, we require that the evolution during the relevant time
period satisfies the density perturbation constraint, which
can be written in the form
$$H^2 / \dot \phi \le 10 \, \delta \ . \eqno(1.7)$$
Second, we assume that during the early stages of inflation,
the evolution of the field $\phi$ is overdamped so that
the $\ddot \phi$ term of Eq.~[1.1] is negligible (along with the
$\Gamma \dot \phi$ term).  This assumption leads to the
simplified equation of motion
$$3 H {d \phi \over d t} = - {d V \over d \phi} \ \ .
     \eqno(1.8)$$
The consistency of neglecting the $\ddot \phi$ term implies
a constraint on the potential of the form
$$\left| {d \over dt} \left( {1 \over 3H} {dV \over d\phi}
     \right) \right| \le \left| {dV \over d\phi} \right| \ \
     ,\eqno(1.9)$$
which we refer to as the overdamping constraint.  This constraint
is often called the ``slowly-rolling'' condition, but we follow
Ref. [2] and avoid this phrase because it suggests a constraint
on $\dot \phi$ (see Eq. [1.10] below) rather than $\ddot \phi$.
Notice that this constraint is a {\it necessary} but not a
{\it sufficient} condition for the $\ddot \phi$ term to be
neglected. Third, we also require that the $\phi$ field
rolls slowly enough that its kinetic energy
contribution to the energy density
of the universe is small compared to that of the vacuum.
Thus, the following constraint must be satisfied during
the inflationary period:
$${ 1 \over 2 } {\dot \phi}^2 \le V_{TOT} \, , \eqno(1.10)$$
where $V_{TOT}$ is the {\it total} vacuum energy density of
the universe.  Notice that additional fields (i.e., in addition
to the inflation field $\phi$) can be present during the
inflationary epoch.  Thus, the total vacuum energy density
$V_{TOT}$ can, in general, include contributions from other
scalar field potentials in addition to $V(\phi)$.
Notice also that the constraint [1.10] was not explicitly
used in our previous work [2].

We have introduced several different potentials and energy
scales and it is important to maintain the distinctions
between them.  The quantity $V(\phi)$ is the potential of
the inflationary field $\phi$ and varies with time
as $\phi$ evolves.  The quantity $V_{TOT}$ is the total
vacuum energy density of the universe and also varies
with time.  The quantity $\Delta V$ is the {\it change} in
the potential $V(\phi)$ over the portion of inflation when
cosmological perturbations are produced; thus, $\Delta V$ is a
given constant for a given inflationary scenario.  Finally, we
have defined $\Lambda$ to be the energy scale of inflation when
cosmological perturbations are produced; to be specific, we define
$$\Lambda^4 \equiv V_{TOT} \Big|_{60} \, , \eqno(1.11)$$
where the right hand side denotes that $V_{TOT}$ is
evaluated when the present-day horizon scale left
the horizon during inflation (this event generally
occurs about 60 e-foldings before the end of inflation).

This paper is organized as follows. In Sec.~II we formulate the
problem for what we call standard inflationary models: models
involving any number of scalar fields that are minimally coupled to
gravity, and that obey the density perturbation and overdamping
constraints.  We define notation and transform the
problem into a mathematically convenient form.  In Sec.~III we
reconstruct the inflationary potential for the case in which the total
primordial spectrum of density perturbations (both scalar and tensor
contributions) is a power-law.  We also show how constraints on the
primordial spectrum lead to corresponding constraints on the
potential.  In Sec. IV, we use our formulation to derive several
additional constraints on the inflationary potential.  In particular,
we constrain $\Delta \phi$ and $\Delta V$ individually; we also derive
a relationship between the width $\Delta \phi$ and the average ratio
of the amplitude of tensor perturbations to scalar perturbations.
We derive further constraints on the fine-tuning parameter $\lft$ in
Sec. V; we show that $\lft$ is also bounded from below and we show the
effects of non-scale-invariant spectra of density perturbations.
Finally, we conclude in Sec.~VI with a summary and discussion of our
results.

\goodbreak \bigskip
\centerline{\bf II. FORMULATION OF THE PROBLEM}
\nobreak \medskip

In this paper, we derive a set of bounds on the properties
of the scalar field potential for a fairly general class of
inflationary scenarios which utilize a slowly rolling field
$\phi$.  We consider inflationary scenarios involving an arbitrary
number of scalar fields that are minimally
coupled to gravity, and which satisfy the density perturbation
constraint of Eq.~[1.7], the overdamping constraint of Eq.~[1.9],
and the inflation constraint of Eq.~[1.10].  We assume that these
constraints hold for a period of $\ntot$ e-foldings, given by
$$N = \int \, H \, dt \, , $$
where the limits of integration correspond to some portion of
the inflationary epoch.  As discussed below, we generally take
$N = 8$ and hence $N$ is not the total number $(\sim 60)$ of
e-foldings required for successful inflation.

The density perturbation constraint must apply for physical
size scales (at the present epoch) in the range 3000 Mpc
(the horizon size) down to about 1 Mpc (the size scale
corresponding to a galactic mass).  This range spans a factor of
3000 in physical size and corresponds to $\ntot$ = $\log[3000]$
$\approx$ 8 e-foldings of the inflationary epoch [8].
Thus, the density perturbation constraint only applies for
about $N=8$ e-foldings.  In this paper, we are mostly
interested in the structure producing portion of the
entire inflationary epoch and we will take $N=8$ as
our ``standard'' value.

The relevant time variable for an inflationary epoch is the
number of e-foldings since the beginning of the epoch.
We therefore adopt a new time variable $x$ defined by
$$d x \equiv {H dt \over \ntot} \ .  \eqno(2.1)$$
The variable $x$ thus ranges from 0 to 1 during the relevant time
period. The point $x=0$ corresponds to the time during inflation
when perturbations on the physical size scale of the horizon
at the present epoch (i.e., 3000 Mpc) were produced.  Keep in
mind that many additional e-foldings of the scale factor could
have taken place before $x=0$.
We also introduce the notation
$$F(x) \equiv - {dV \over d\phi} \ \ ,\eqno(2.2)$$
where $F$ represents a force.

In this newly defined notation, the overdamping constraint is
written as
$$\left| H {d \over dx} \left({F \over H}\right)\right| \le 3
     \ntot F \ , \eqno(2.3)$$
and the density perturbation constraint is
$$3 H^3 / F \le 10 \, \delta \ .\eqno(2.4)$$
Using the equation of motion [1.8] and the relation
$$H^2 = (8 \pi/3) V_{TOT} /\mp^2 \ , $$
we can write the condition
that the universe is dominated by potential energy rather
than kinetic energy (Eq. [1.10]) in the form
$$F < \Bigl( {27 \over 4 \pi} \Bigr)^{1/2} \,
H^2 \, \mp \, , \eqno(2.5)$$
where $\mp$ is the Planck mass.
Furthermore, the quantities $\Delta V$ and $\Delta \phi$ can be
written in the form
$$\Delta V = {\ntot \over 3} \int_0^1 (F^2 / H^2) \, dx
     \eqno(2.6)$$
$$\Delta \phi = {\ntot \over 3} \int_0^1 (F / H^2) \, dx
     \ . \eqno(2.7)$$
We have chosen our sign convention so that $\Delta V$ is a
positive quantity and so that $x=0$ at the beginning of the
constrained time period.  Keep in mind that $\Delta V$ and
$\Delta \phi$ are the changes in the potential and the
inflation field during the $N=8$ e-foldings during which
cosmic structure is produced; they are {\it not} the total
changes in these quantities over the entire inflationary epoch.

In Sections IV and V, we find general constraints on the inflationary
potential.  For these calculations, we introduce the formulation
described below.  For the sake of definiteness, in Sections IV and V
we assume that the density perturbation constraint of equation [2.4]
is saturated at the epoch $x=0$, i.e., when the present-day horizon
scale left the horizon during inflation (this assumption and the
following definitions are {\it not} used in the reconstruction of the
potential in Sec. III).  Physically, this assumption means that scalar
density perturbations are responsible for the observed fluctuations in
the cosmic microwave background as measured by the COBE satellite.
We thus have
$${3 H_B^3 \over F_B } = 10 \delta \, , \eqno(2.8{\rm a})$$
where the subscripts denote the epoch at which $x=0$ and where we
consider $\delta$ to be a known number ($\sim 2 \times 10^{-5}$).
Equation [2.8a] will be used in Sections IV and V.

We note that in general tensor perturbations
may produce some fraction of the total
perturbations; in this case, one should replace $\delta$ in equation
[2.8a] by the corresponding smaller value $\delta_S$ which denotes
only the scalar contribution, i.e.,
$${3 H_B^3 \over F_B } = 10 \delta_S \, . \eqno(2.8{\rm b})$$
In this case, the general form of the results derived in Sections
IV and V remain the same with $\delta$ replaced by $\delta_S$.
For completeness, we also note that the maximum of the density
perturbation constraint need not occur at $x=0$; this complication
is considered in Ref. [2] and will not significantly affect the
results of this paper.

For convenience, we rescale the functions $F$ and $H$ by their
starting values in order to obtain dimensionless quantities, i.e.,
$$\eqalignno{f(x) &\equiv
F(x) / F_B \ , &(2.9{\rm a})\cr
  \noalign{\smallskip}
  h(x) &\equiv {H (x) / H_B} \ . &(2.9{\rm b})\cr}$$
Since we can use equation [2.8] to eliminate $F_B$ from our
equations, we are left with a single unknown parameter --
namely $H_B$.  We choose to eliminate $H_B$ in favor of the
energy scale $\Lambda$ at $x=0$, i.e., we define
$$H_B^2 \equiv {8 \pi \over 3} {\Lambda^4 \over \mp^2}
\, = {8 \pi \over 3} {V_{TOT} (x=0) \over \mp^2} \, . \eqno(2.10)$$
The quantity $\Lambda^4$ is equal to the value of the total
vacuum energy density of the universe at $x=0$ (which occurs
$\sim$60 e-foldings before the end of inflation).

For mathematical convenience, we also eliminate $f(x)$ in
favor of a new function $p(x)$ defined by
$$p(x) \equiv \sqrt{f(x) \over h(x) } \ \ . \eqno(2.11)$$
In terms of the dimensionless functions $p(x)$ and $h(x)$,
the physical quantities of interest in this paper can be
written
$$\Delta V = {16 \pi^2 \over 75} {N \over \delta^2} \,
{\Lambda^8 \over \mp^4} \, \int_0^1 p^4 \, dx \, , \eqno(2.12)$$
$$\Delta \phi = \Bigl( {2 \pi \over 75} \Bigr)^{1/2}
{N \over \delta} \, {\Lambda^2 \over \mp} \,
\int_0^1 (p^2/h) \, dx \, , \eqno(2.13)$$
$$\lft = {300 \delta^2 \over \ntot^3} \,
\, \jfun [p,h] \ , \eqno(2.14)$$
where $\jfun$ is the functional defined by
$$\jfun[p,h] = {\int_0^1 p^4 \, dx \over \Bigl[ \int_0^1
     (p^2/h) \, dx \Bigr]^4 } \ \ . \eqno(2.15)$$
In terms of the new functions $p$ and $h$, the density
perturbation constraint [2.4] can be written as
$$p(x) \ge h(x) \eqno(2.16)$$
and the overdamping constraint as
$$\Biggl| {1 \over p} {dp \over dx} \Biggr| \le {3 \over 2}
     \ntot \ .\eqno(2.17)$$
Using equations [2.8 -- 2.11], the constraint of
equation [2.5] becomes
$${p^2 \over h} \le {15 \over 2 \sqrt{2} \pi} \, \delta
\, {\mp^2 \over \Lambda^2} \, \equiv \beta \, , \eqno(2.18)$$
where we have defined the right hand side of the
inequality to be the dimensionless parameter $\beta$.
In addition, the functions $p$ and $h$ are subject to the
initial conditions
$$p(0) = 1 = h(0) \ .\eqno(2.19)$$
Notice that the ratio $h/p$ is constrained to have a maximum
value of unity, but the individual functions $h$ and $p$ can
vary substantially.

This concludes the formulation of the problem.
We want to bound the physical quantities defined
by equations [2.12 -- 2.15] subject to the constraints
of equations [2.16 -- 2.18] and the initial conditions
[2.19].

\goodbreak
\bigskip
\centerline{\bf III. RECONSTRUCTION OF INFLATIONARY POTENTIALS}
\medskip

In this section, we consider the problem of reconstructing the scalar
field potential.  As noted by many authors in the recent literature
[15], knowledge of both the scalar perturbations and the tensor
perturbations allows one to reconstruct a portion of the scalar field
potential that gives rise to inflation.  In this paper, we formulate
the reconstruction problem in terms of the variables defined in the
previous section.  We then find a semi-analytic solution for the potential
for the case of (total) perturbation spectra which are pure power-laws
(see equation [3.12]).

To preview some of the most interesting results of this section,
we refer the reader to Figures 1 -- 5.  There we
show how the shape of the potential depends on the
perturbation spectrum and on the relative contribution of tensor
modes.  For example, when tensor modes
provide a significant fraction of the total, the potentials
$V(\phi)$ are concave upward for all spectral indices considered
in this paper $n = 0.5 - 1$ (e.g., Figure 1).  For the opposite
case where tensor modes provide a negligible contribution to the
total, the potentials are concave downward (e.g., Figure 4).
For this latter case, the potential shape is well approximated
by a cosine (see Figure 5) as in the model of Natural Inflation [7].

For the rest of this section we show how these results
are obtained.  In addition, we comment on their usefulness
for the case when the exact power law index is not known,
but instead there is a range consistent with
the existing status of observations.
Although our knowledge of the true primordial
spectrum of perturbations is not exact, constraints may be
placed on the spectrum; we show how {constraints} on the primordial
power spectrum produce corresponding constraints on the scalar field
potential.

The scalar perturbations (see equation [1.2]) will
generally be some function of the variable $x$
introduced in the previous section, i.e.,
$${\delta \rho \over \rho} \Bigg|_{\rm scalar} \, =
{1 \over 10} {H^2 \over \big| {\dot \phi} \big| } \,
\equiv \, \delta_S (x) \, . \eqno(3.1)$$
Similarly, the tensor perturbations (gravity wave
perturbations) can be written in the form
$${\delta \rho \over \rho} \Bigg|_{\rm GW} \, =
{ \sqrt{32 \pi} \over 30} \, {H \over \mp}
\equiv \, \gwper (x) \, ,  \eqno(3.2)$$
where the right hand side is some function of $x$.
We note that the expressions used here are correct only
to leading order in the ``slow-roll'' approximation.
Although higher order corrections to these expressions
have been calculated [20], the leading order terms
are adequate for our purposes.

For the particular case in which inflation arises from
a single scalar field $\phi$ with a potential $V (\phi)$,
we can write the above expressions in terms of the
potential.  The tensor modes are related to the potential
through equation [3.2], which can be written as the expression
$$V(x) = {675 \over 64 \pi^2} \mp^4 \gwper^2 (x) \, .
\eqno(3.3)$$
Similarly, the scalar modes are related to the potential
through equation [3.1], which can be written in the form
$$-{1 \over V^2} {dV \over dx} = {16 \pi^2 N \over 75}
\mp^{-4} \delta_S^{-2} (x) \, . \eqno(3.4)$$
The combination of these two equations thus implies
the following simple differential equation
$$- {1 \over \gwper^3} {d \gwper \over dx} =
{9 N \over 8} \delta_S^{-2} \equiv C \delta_S^{-2}
\, , \eqno(3.5)$$
where we have defined the constant $C$ = $9N/8$ (see Ref. [25]).

The two types of perturbations add in quadrature,
so that the total spectrum of primordial perturbations,
which we denote as $q(x)$, can be written as the sum
$$q^2 (x)  = \delta_S^2 (x) \, + \, \gwper^2 (x) \, .
\eqno(3.6)$$
If we assume that the total spectrum $q(x)$ is a known function,
we can then combine the equations [3.5] and [3.6] to obtain
a single differential equation for $\gwper$:
$$\bigl[ \gwper^2 - q^2 \bigr] \, {d \gwper \over dx} =
C \gwper^3 \, . \eqno(3.7)$$
Thus, if the primordial spectrum $q(x)$ were known exactly, we could
simply solve the above differential equation for $\gwper(x)$ and then
solve for the scalar field potential $V(x)$ [21].  Notice that we must
also specify the initial condition $\gwper(0)$, i.e., the amplitude of
the tensor modes at $x=0$.  Since $\phi(x)$ is directly calculable from
the equation of motion once we know $V(x)$, the usual form of the
potential $V(\phi)$ as a function of the scalar field can also be
obtained.  This hypothetical ``solution'' for the potential is
correct to leading order in the ``slow roll'' approximation
(see also Ref. [15]).

One problem with the above discussion is that we do not know the
true primordial spectrum $q(x)$.  However, the total spectrum of
perturbations is often assumed to be a power-law in wavenumber $k$,
i.e., the amplitudes of the perturbations vary
with physical length scale $L$ according to the law
$$q \, \sim \, {\delta \rho \over \rho} \Bigg|_{\rm hor}
\sim L^{(1-n)/2} \, , \eqno(3.8)$$
where the subscript denotes that $\delta \rho/\rho$
is evaluated at the time of horizon crossing.  The parameter
$n$ is the power-law index of the primordial power spectrum,
$$P(k) \sim \big| \delta_k \big|^2 \sim k^n \, , \eqno(3.9)$$
and $k$ is the wavenumber of the perturbation [22].
Notice that the left-hand-side of equation [3.8] is to
be evaluated when the perturbation of lengthscale $L$
enters the horizon.  Notice also that $n=1$ corresponds
to a scale-invariant spectrum and that $n < 1$ corresponds
to spectra with more power on large length scales.
We stress that a considerable amount of processing
in required to convert the primordial spectrum into
observable quantities and such work is now being
vigorously pursued [26]; this transformation between the
primordial spectrum and actual observed quantities
is generally very complicated and model dependent.

For now we take the exponent $n$ as given and proceed
to a reconstruction of the potential.  Subsequently
we will consider the situation where the primordial
spectrum is not entirely known and not necessarily
pure power law.  We define a new function
$$\vr (x) \equiv \gwper^2 / q^2 \, , \eqno(3.10)$$
where $\vr (x) \le 1$ by definition.  For convenience
we also define $\alpha = N(1-n)/2$.  Thus for the scale
invariant case of $n = 1$ we have $\alpha = 0$,
while for $n < 1$ we have $\alpha > 0$.
Notice that for the
case in which $\alpha = 0$ (corresponding to a scale-invariant
perturbation spectrum), the function $\vr$ is proportional to
the potential $V$ (see equation [3.3]). In terms of this
new function $\vr$, the differential equation [3.7] becomes
$${(\vr - 1) \over 2 \vr} {d\vr \over dx} =
C \vr + \index (\vr - 1) \, . \eqno(3.11)$$
For the case of $\index = constant$,  equation [3.11]
can be integrated to obtain the solution
$${-C \over C + \index} \, \log \Biggl[
{ (C + \index) \vr - \index \over (C + \index) \vr_B - \index }
\Biggl] \, + \, \log(\vr / \vr_B) = 2 \index x \, , \eqno(3.12)$$
where $\vr_B$ denotes the function $v(x)$ evaluated at $x=0$.
Keep in mind that $\vr_B$ represents the ratio (squared)
of the amplitude of tensor modes to the total amplitude of
density fluctutations.

We can use the above results to reconstruct the inflationary
potential as follows. Once the initial condition (i.e., $\vr_B$)
is specified, equation [3.12] provides an implicit, but analytic,
solution for $\vr (x)$.  We can then use equation [3.10] to find
$\gwper^2$ and then use equation [3.3] to find the potential as
a function of $x$. Notice that we have found $V(x)$ and not
$V(\phi)$.  In order to make this conversion, we must also solve
the equation of motion for $\phi(x)$; this equation is written
in integral form in equation [2.7].

We have performed the reconstruction process outlined above for
varying values of the initial ratio $\vr_B$ and for varying choices
of the index $n$.  The results are shown in Figures 1 -- 4.  For each
choice of $\vr_B$ (which determines the relative amplitude of the
tensor modes), the figures show the resulting potentials $V(\phi)$ for
$n$=0.5, 0.6, 0.7, 0.8, 0.9, and 1.0. The open symbols represent
$x=1$, i.e., the epoch at which galaxy sized perturbations left the
horizon during inflation.  Keep in mind that this reconstruction
process only contains information about the potential during the
$N=8$ e-foldings when structure-forming perturbations are produced.
This procedure says nothing about the potential at subsequent epochs.

The results shown in Figures 1 -- 4 show interesting general trends.
For the case in which tensor perturbations produce a substantial
contribution to the total (e.g., in Figure 1), the potentials
$V(\phi)$ are concave upward.  For the opposite case in which tensor
modes are negligible (e.g., in Figure 4), the potentials are concave
downward and somewhat reminiscent of a cosine potential.  To follow up
on this latter issue, we fit a cosine potential to the reconstructed
potential for the specific case $n = 0.6$ and $\vr_B$ = $10^{-4}$.
The result is shown in Figure 5.  Thus, for perturbation spectra with
moderate departures from scale invariance and little contribution from
tensor modes, the reconstructed potential looks very much like a
cosine.  This type of potential is used in the model of Natural
Inflation [7] and was first suggested for reasons of technical
naturalness. In particular, the required small parameter $\lft$
(see equations [1.5] and [1.6]) occurs naturally in this model.

As mentioned above, the transformation between actual observed
quantities such as the microwave anisotropy and the primordial
spectrum is complicated and model dependent.  Thus, a definitive
prediction for $q(x)$ may be difficult to obtain in the near future.
However, the observations can be used to imply {constraints} on the
spectrum $q(x)$.  For example, our analysis of the observations may
imply that the true spectrum lies within some range of power laws.
Then we can obtain the range of possibilities for the potential
from the figures by restricting ourselves to those curves
corresponding to that range of power laws.

For example, we might reasonably require that the amplitude
of the perturbations does not change too much with varying
lengthscale (wavenumber).  In the present formulation, this
statement takes the form
$$A \le \index (x)  \le B \, , \eqno(3.13)$$
where we have defined
$$\index (x) \equiv - {1 \over q} {d q \over dx} \, .
\eqno(3.14)$$
For the special case in which the primordial spectrum is a pure
power-law, the index $\index (x)$ defined here is a constant
independent of $x$ and is related to the spectral index $n$
through the identity $\alpha = N (1-n)/2$.  In general, the
index $\index$ will not be constant, but we expect that the
function $\index (x)$ will be a slowly varying function.

Constraints of the form [3.13] imply corresponding constraints
on the potential (for a given set of initial conditions).
If the index $\alpha$ is constrained as in equation [3.13],
then the amplitude $\delta_T$ is constrained to lie between
the solutions found with $\alpha = A$ and $\alpha = B$.
Since the potential is proportional to $\delta_T^2$, the
potential will be similarly constrained (see Figure 1).
In other words, the potential is allowed to be in the range
of curves corresponding to the appropriate range of indices $n$
in the figures.

In this section, we have considered the reconstruction of
the inflationary potential.  Building on previous work by
several groups [15], we found a semi-analytic solution
for the potential for pure power-law spectra, and
plotted our results in Figures 1 -- 5.

\goodbreak
\bigskip
\centerline{\bf IV. CONSTRAINTS ON THE HEIGHT AND WIDTH}
\centerline{\bf OF INFLATIONARY POTENTIALS}
\medskip

In this section we present a series of constraints on the scalar
field potential.  These bounds apply to all inflationary models
which belong to the general class of models defined in Sec. II,
i.e., models which obey the density perturbation constraint [2.16],
the overdamping constraint [2.17], and the condition of vacuum
energy domination [2.18].

\bigskip
\centerline{\bf A. Relationship Between the Width of the Potential}
\centerline{\bf and the Relative Amplitude of Tensor Perturbations}
\nobreak \medskip

Tensor perturbations (i.e., gravity wave perturbations) can
arise during the inflationary epoch.  Many authors have explored
their effects [12]. The ratio of the amplitude of tensor
perturbations to the scalar perturbations can be written
in the form
$$\rat = { \sqrt{32 \pi} \over 3} \,
{ \big| {\dot \phi} \big| \over \mp H} \, . \eqno(4.1)$$
In terms of the functions defined in Sec. II of this paper,
this ratio can be expressed in the form
$$\rat (x) = {8 \pi \over 15 \sqrt{3} } \,
{\Lambda^2 \over \delta \mp^2} \, {p^2 \over h}
\, . \eqno(4.2)$$
If we now take the average value of $\rat$ over the portion
of inflation when structure is produced, we obtain
$$\rave = {8 \pi \over 15 \sqrt{3} } \,
{\Lambda^2 \over \delta \mp^2} \,
\int_0^1 (p^2 / h) \, dx \, \, , \eqno(4.3)$$
where $\rave$ denotes the average value.

We must now derive an expression for the width of the potential
$\Delta \phi$ during the $N \approx 8$ e-foldings when cosmological
structure can be produced.  Using the functions defined in Sec. II
and equation [2.13], we can write the width $\Delta \phi$ as
$$\Delta \phi = \Bigl( {2 \pi \over 75} \Bigr)^{1/2}
{N \over \delta} \, {\Lambda^2 \over \mp} \,
\kfun [p, h] \, , \eqno(4.4)$$
where $\kfun$ is the functional defined by
$$\kfun [p, h] = \int_0^1 (p^2/h) \, dx . \eqno(4.5)$$

Comparing equation [4.3] for the ratio $\rave$ with equation [4.4]
for $\Delta \phi$, we discover the simple relationship
$$\rave = { \sqrt{32 \pi} \over 3 N} \,  \,
{\Delta \phi \over \mp} \, . \eqno(4.6)$$
Thus, for the $N \approx 8$ e-foldings where density fluctuations
of cosmological interest can be produced, we obtain
$$\rave \approx 0.42 \Delta \phi/\mp \, . \eqno(4.7)$$
One important implication of this result can be stated as follows:
If tensor perturbations play a major role in the formation of structure,
then the width $\Delta \phi$ during the appropriate part of inflation
must be comparable to the Planck mass $\mp$.  [Notice that since this
argument applies only to the average value of the ratio $\rat$, it is
logically possible for tensor modes to be significant at some
particular length scale, even though the average $\rave$ is small.
However, the overdamping constraint prevents the potential (and hence
$\rat$) from changing very quickly and all but eliminates this
possibility.]

As we show below, the allowed width $\Delta \phi$ is
also bounded from above (see equation [4.13]). Thus, we
also obtain an upper limit on the ratio $\rave$, i.e.,
$$\rave \le 1.6 \, . \eqno(4.8)$$
Although scalar perturbations can be larger than
tensor modes by an arbitrarily large factor, the converse
is not true: tensor modes can be {\it at most} a factor of
$\sim$1.6 larger than the scalar contribution.

\bigskip
\centerline{\bf B. Constraints on the Width of the Potential}
\nobreak
\medskip

In this section we find both lower (equation [4.11]) and upper
(equation [4.13]) bounds on the width of the potential.  These
bounds apply to the portion of the inflationary epoch when
cosmologically interesting perturbations are produced.

In order to find bounds on $\Delta \phi$, we must find bounds
on the functional $\kfun$ in equation [4.5].  This functional
can be minimized by saturating the density perturbation
constraint [2.16]; we thus obtain
$$\kfun \ge \int_0^1 p \, dx \, . \eqno(4.9)$$
A lower limit on the remaining integral can be found by
using the overdamping constraint [2.17].  Since we have
fixed $p(0)$ and we want to find the smallest possible value
for the integral in equation [4.9], we must choose
the sign of the derivative $dp/dx$ to be negative and as
large as possible given the overdamping constraint.
Our limit thus becomes
$$\kfun \ge {2 \over 3N} \Bigl\{ 1 - {\rm e}^{-3N/2} \Bigr\}
\approx {2 \over 3N} \, , \eqno(4.10)$$
where the last approximate equality has a relative error
less than $10^{-5}$ for $N=8$.  As we show in Appendix A,
the lower bound of equation [4.10] is in fact the greatest
lower bound for this problem.
Putting the above results together, we obtain
$${\Delta \phi \over \mp}
\ge {2 \over 15} \Bigl( {2 \pi \over 3} \Bigr)^{1/2}
\delta^{-1} {\Lambda^2 \over \mp^2} \, \, ; \eqno(4.11)$$
this is the desired lower bound.

We now derive an upper limit on the allowed width $\Delta \phi$.
In this case, we use the constraint of equation [2.18] which
implies that the kinetic energy of the rolling field does not
dominate the vacuum energy density.  Using this constraint,
we see immediately that the functional $\kfun$ of equation
[4.5] is bounded by
$$\kfun [p, h] \le \beta \, , \eqno(4.12)$$
where $\beta$ is the dimensionless parameter defined in
equation [2.18].  Combining this bound with expression
[4.4] for $\Delta \phi$, we find
$${\Delta \phi \over \mp}
\le N \sqrt{ {3 \over 4 \pi} } \, . \eqno(4.13)$$
For the standard choice $N=8$, this limit implies
$\Delta \phi/ \mp \le$ 3.9. We note that this upper limit follows
directly from the definition of $\Delta \phi$ and the
condition [1.10] which must be met in order for inflation
to take place.  In particular, this bound is independent
of the density perturbation constraint.

Putting all of the results of this subsection together,
we find that the change $\Delta \phi$ in the field
is constrained to lie in the range
$${2 \over 15} \Bigl( {2 \pi \over 3} \Bigr)^{1/2}
\delta^{-1} {\Lambda^2 \over \mp^2} \, \le \,
{\Delta \phi \over \mp} \,
\le \, N \sqrt{ {3 \over 4 \pi} } \, . \eqno(4.14)$$
Another way to write this constraint is in terms of
the Hubble parameter $H_B$ at the epoch $x=0$, i.e.,
$${H_B \over 15 \delta} \, \le \, \Delta \phi \,
\le \, N \sqrt{ {3 \over 4 \pi} } \, \mp \, . \eqno(4.15)$$
The above bounds suggest that the change $\Delta \phi$
in the scalar field is rather constrained.  For all
cases, $\Delta \phi$ cannot be much larger than the
Planck scale $\mp$.  The lower bound shows that the
change in the field $\Delta \phi$ must be at least
a factor of $\sim 10^3$ larger than the Hubble parameter.
Notice that for an inflationary energy scale $\Lambda$
comparable to the GUT scale, the change $\Delta \phi$
in the inflation field must be larger than $\sim \mp$.
In the following section, we calculate the width $\Delta \phi$
for three ``standard'' inflationary potentials and show that
the condition $\Delta \phi \sim \mp$ is in fact typical.

\bigskip
\goodbreak
\centerline{\bf C. Width of the Potential for Examples}
\medskip

In this section, we calculate the width $\Delta \phi$ for
several standard inflationary models, including monomial potentials
(such as in the original version of
Chaotic Inflation [6]), exponential potentials, and
cosine potentials (such as in Natural Inflation [7]). Here, we
write the number $N$ of e-foldings as
$$N = \int H \, dt \, = {8 \pi \over \mp^2} \,
\int {V \, d\phi \over \Big| {d V/ d\phi} \Big| }
\, , \eqno(4.16)$$
where we have used the slowly rolling version of the
equation of motion to obtain the second equality.
In the integral in equation [4.16], the range of
integration corresponds to the range $\Delta \phi$
of interest.  For the cases of monomial potentials
and exponential potentials, we only consider the portion
of inflation during which density fluctuations of
cosmologically interesting sizes are produced
(i.e., we take $N = 8$ as usual).
For the case of Natural Inflation (cosine potentials),
we consider the entire overdamped phase of inflation
(i.e., we take $N \approx 60$ or so).

We first consider the monomial potential
of the form
$$V(\phi) = \lambda_j \phi^j \, , \eqno(4.17)$$
where $j$ is an integer.  For this class of models,
the number of e-foldings is given by
$$N = {4 \pi \over j} \mp^{-2} \, \Bigl[
\phi_1^2 - \phi_2^2 \Bigr] \, , \eqno(4.18)$$
where $\phi_1$ and $\phi_2$ are the initial and
final values of the field.  Without loss of generality,
we take $\phi_1 > \phi_2$ [27]. Solving for the
width $\Delta \phi$, we find
$${\Delta \phi \over \mp}  = {\phi_1 - \phi_2 \over \mp}
= \Bigl[ N j / 4 \pi + (\phi_2/\mp)^2
\Bigr]^{1/2} -  \phi_2/\mp \, , \eqno(4.19)$$
where we have eliminated $\phi_1$ using equation [4.18].
We examine this expression in two limits, $\phi_2 \ll \mp$
and $\phi_2 \ge \mp$.  In the first case,
$\Delta \phi/\mp \sim (Nj/4\pi)^{1/2} \sim 1$ for $N=8$.
The second possibility is that the final value
of the field is in the regime $\phi_2 \ge \mp$.  Thus, either
$\Delta \phi$ is comparable to the Planck scale or $\phi_2$
is larger than the Planck scale.  In either case,
an energy scale comparable to or larger than the Planck
scale must be present in the problem.

As the next example, we consider an exponential potential
of the form
$$V(\phi) = V_0 \exp [ - \phi/\sigma ] \, , \eqno(4.20)$$
where $\sigma$ is the energy scale that characterizes the
fall-off of the potential.  We note that this form is often
used as an approximation to the true potential and is valid
for only part of the inflationary epoch.  However, as long
as the form [4.20] holds for a few e-foldings of the scale
factor, the following argument is valid.
Using the definition [4.16], we obtain
$$N = 8 \pi {\Delta \phi \over \mp} {\sigma \over \mp}
\, . \eqno(4.21)$$
Solving for $\Delta \phi$, we find
$${\Delta \phi \over \mp} = {N \over 8 \pi}
{\mp \over \sigma} \, \approx 0.32
{\mp \over \sigma} \, , \eqno(4.22)$$
where we have used $N = 8$ to obtain the final approximate equality.
Equation [4.22] shows that either the width $\Delta \phi$ must be
comparable to the Planck scale $\mp$, or, the fall-off scale $\sigma$
must be much larger than $\mp$.  Once again, an energy scale
comparable to or larger than the Planck scale must be present.

Finally we consider a cosine potential, i.e.,
$$V(\phi) = \Lambda^4 \bigl[ 1 + \cos (\phi/f) \bigr]
\, , \eqno(4.23)$$
such as that found in Natural Inflation [7].
For this case, we find
$$N_{TOT} = {16 \pi f^2 \over \mp^2} \,
\log \Biggl\{ {\sin (\phi_2/2f) \over  \sin (\phi_1/2f) }
\Biggr\} \, ,  \eqno(4.24)$$
where we have denoted the number of e-foldings as $N_{TOT}$
$\approx 60$ to emphasize that we do not use $N=8$ for this case.
This potential has a definite width, namely $f$.
Thus, in this case, we have
$$\Delta \phi_{TOT} \sim f \sim \mp \Bigl(
{N_{TOT} \over 16 \pi} \Bigr)^{1/2}
\Biggl\{ \log [2f /\phi_1] \Biggr\}^{-1/2} \, , \eqno(4.25)$$
where $\Delta \phi_{TOT}$ is the width of the potential
over 60 e-foldings (rather than merely the 8 of structure formation).
In this equation we have used the fact that $\phi_2 \sim f$
(more precisely, we assume that $\log[\sin(\phi_2/2f)]$ is
of order unity). Thus, unless the remaining logarithmic factor
in equation [4.25] becomes very far from unity, this potential has
a width which is of order the Planck scale $\mp$.  A detailed
treatment of the conditions for sufficient inflation with
this potential (see Ref. [7]) confirms that $f$ and
$\Delta \phi$ must be near the Planck scale $\mp$ for
this model.

We thus conclude that for these particular examples, the inflationary
potentials contain energy scales which are comparable to (or larger
than) the Planck scale $\mp$.  While all of the models considered here
have $\Delta \phi \sim \mp$, we note that constraint of equation
[4.14] is much less restrictive for small values of $\Lambda$ (the
energy scale of inflation); for example, if $\Lambda = 10^{12}$ GeV,
the bound becomes very weak, $\Delta \phi \ge 10^{-10} \mp$.
This apparent discrepancy is easy to understand.  For simple
``well-behaved'' potentials (such as the examples considered here),
the integral in equation [4.16] $\sim (\Delta \phi)^2$, and equation
[4.16] reduces to
$$N \sim  8 \pi {(\Delta \phi)^2 \over \mp^2} \,
\, .  \eqno(4.26)$$
We thus naively expect that any sufficiently well-behaved potential
will have $\Delta \phi \sim \mp$.  However, in the general bound of
equation [4.14], we allow the potential to take {\it any} form,
provided only that the density perturbation constraint and the
overdamping constraint are satisfied.  This considerable extra
freedom leads to the appreciably weaker bound.

\bigskip
\goodbreak
\centerline{\bf D. Constraints on the Change in Height of the Potential}
\medskip

In this section, we constrain the allowed change in height
$\Delta V$ of the potential during the overdamped phase of
inflation when cosmological structure may be produced.
We find both upper and lower bounds on $\Delta V$ and
show that these results imply an upper limit on the energy
scale of inflation (during this phase).

The change in potential $\Delta V$ is given by equation [2.12].
In order to constrain $\Delta V$, we must constrain the
functional
$$\lfun [p] = \int_0^1 p^4 \, dx \, ,  \eqno(4.27)$$
subject to the constraints of equations [2.16 -- 2.19].
Perhaps counterintuitively, the upper limit on the energy
scale $\Lambda$ arises from a lower limit on the change in
height $\Delta V$.  We thus consider this limit first.
Using the overdamping constraint [2.17], we find
$$\lfun \ge {1 \over 6N} \Bigl\{ 1 - {\rm e}^{-6N}
\Bigr\} \approx {1 \over 6N} \, . \eqno(4.28)$$
We note that this bound is the greatest lower bound
for this constraint problem.  Putting in the dimensionful
quantities, we obtain the limit
$$\Delta V \ge {8 \pi^2 \over 225} \delta^{-2} \mp^{-4}
\Lambda^8 \, . \eqno(4.29)$$
Since the vacuum contribution to the energy density cannot
become negative [28], we must also require
$$\Lambda^4 \ge \Delta V \, . \eqno(4.30)$$
Combining these two limits and solving for the
scale $\Lambda$, we obtain the desired limit on the
energy scale $\Lambda$,
$${\Lambda \over \mp } \le
\delta^{1/2} \Bigl( {15 \sqrt{2} \over 4 \pi}
\Bigr)^{1/2} \, \approx 6 \times 10^{-3} . \eqno(4.31)$$
This constraint implies that the structure producing portion
of inflation must take place at an energy scale roughly
comparable to (or less than ) the GUT scale.
Although found by slightly different methods, this constraint
is equivalent to that derived earlier by Lyth [16]. Notice
also that this constraint is comparable to that obtained in
Ref. [19] by requiring that tensor perturbations are not in
conflict with the COBE measurement.  Finally, we note that
this constraint is equivalent to the requirement $\beta \ge 1$.

We note that an upper bound for $\Delta V$ also exists.
This bound can be obtained by finding an upper bound for
the functional $\lfun$ (see Eq. [4.27]).  Using the overdamping
constraint with the opposite sign, we find
$$\lfun [ p ] \le {1 \over 6N} \Bigl\{ {\rm e}^{6N} - 1
\Bigr\} \approx {{\rm e}^{6N} \over 6N} \, , \eqno(4.32)$$
and the corresponding bound on $\Delta V$ becomes
$$\Delta V \le {8 \pi^2 {\rm e}^{6N} \over 225} \,
\delta^{-2} \mp^{-4} \Lambda^8 \, . \eqno(4.33)$$
At first glance, the bound of equation [4.33] may
not seem significant.  However, at sufficiently small
energy scales $\Lambda$, this bound becomes very
severe.  Let us define a new dimensionless parameter
$$\eta \equiv {\Delta V \over V_B} \, \le \,
{8 \pi^2 {\rm e}^{6N} \over 225} \,
\delta^{-2} \Bigl( {\Lambda \over \mp} \Bigr)^4 \sim 6 \times
10^{29} \Bigl( {\Lambda \over \mp} \Bigr)^4
\, , \eqno(4.34)$$
where $V_B = \Lambda^4$ is the value of the total vacuum energy
density at $x=0$.  For example, if we consider inflationary
models at low energies such as $\Lambda = 1$ TeV,
we obtain the bound $\eta \le 10^{-34}$.

It is easy to see why the above result makes inflation at low
energies problematic.  During the $N=8$ e-foldings of inflation
where structure is produced, we must have $\eta = \Delta V/V_B
\ll 1$.  However, the vacuum energy density must be essentially
zero at the end of the entire inflationary epoch; thus,
during the following 52 e-foldings of inflation, we must
have $\Delta V/V_B$ $\sim$ 1 [29].  It seems unlikely that
particle physics models will produce a scalar field
potential with such extreme curvature.

Before leaving this section, we note that the above argument defines
a suggestive lower bound for the energy scale of inflation.  Arguing
{\it very} roughly, we expect that models of inflation with the
parameter $\eta$ very much smaller than unity are
difficult to obtain. As shown
by equation [4.34], the parameter $\eta$ decreases with the energy
scale $\Lambda$ of inflation.  We thus obtain a suggestive lower bound
for $\Lambda$ by requiring that $\eta$ be larger than some ``not too
unnaturally small number'', say 1/10.  The requirement that $\eta >
1/10$ implies that the energy scale of inflation to obey the constraint
$$\Lambda \, \ge \, \mp \, \delta^{1/2} \, {\rm e}^{-3N/2} \,
\Bigl[ {2250 \over 8 \pi^2} \Bigr]^{1/4} \,
\approx 10^{-7} \, \mp \, \approx 10^{12} {\rm GeV} \, ,
\eqno(4.35)$$
where the numerical value was obtained using $N=8$.  We stress that
this bound is {\it not} a firm lower limit on the energy scale
$\Lambda$, but it is suggestive.  In particular, for energy scales
$\Lambda$ much less than about $10^{12}$ GeV, the parameter $\eta$
becomes very small compared to unity.

\bigskip
\centerline{\bf V. CONSTRAINTS ON THE FINE-TUNING PARAMETER}
\medskip

In this section, we constrain the fine-tuning parameter $\lft$
as defined by equation [1.5].  In our previous paper [2], we
found a firm upper limit on the parameter $\lft$.  In this paper,
we first complete the argument by finding a lower limit on $\lft$.
Next, we show how density perturbation spectra which are not
scale-invariant can place slightly tighter bounds on $\lft$.

\bigskip
\centerline{\bf A. Lower Bound on the Fine-Tuning Parameter}
\medskip

In order to bound $\lft$, we need bounds on both the height $\Delta V$
and the width $\Delta \phi$.  We have already shown that $\Delta V$
is bounded from below by equation [4.29] and that $\Delta \phi$ is
bounded from above by equation [4.13].  Combining these two results
thus gives us a lower bound on the ratio $\lft$ =
$\Delta V/(\Delta \phi)^4$, i.e.,
$$\lft \ge {128 \over 2025} \, {\pi^4 \over N^4} \,
\delta^{-2} \, \Bigl( {\Lambda \over \mp} \Bigr)^8
\, . \eqno(5.1)$$
Combining this result with the general bound of
Ref. [2], we find that $\lft$ is confined to the
range
$$ {128 \over 2025} \, {\pi^4 \over N^4} \,
\delta^{-2} \, \Bigl( {\Lambda \over \mp} \Bigr)^8
\, \le \lft \le \, {2025 \over 8} \, \delta^2
\, . \eqno(5.2)$$
For example, if we use representative values of $\Lambda/\mp$
$\sim$ $10^{-3}$, $N$ = 8, and $\delta \sim 2 \times 10^{-5}$,
 the allowed range for the parameter
$\lft$ becomes
$$6 \times 10^{-19} \, \le \, \lft \, \le \, 10^{-7} \, .  \eqno(5.3)$$
The bound on $\Delta \phi$ was obtained independently
of the density perturbation constraint; the bound on $\Delta V$
was obtained by requiring $\delta \rho/\rho \le \delta$.

\bigskip
\centerline{\bf B. Effects of Departures from Scale Invariance}
\nobreak
\medskip

For the limits presented thus far (see Sec. IV), we have used the density
perturbation constraint in the form of equation [1.7], which assumes
that the amplitude of the density perturbations produced by inflation
must be less than a constant value (i.e., the constraint is the same
for all perturbation wavelengths).  However,
one way to explain current cosmological data is with
density perturbations with a non-scale-invariant
spectrum [7, 29].  In this case the departures from scale invariance
imply that our universe has density perturbations which exhibit
{\it more power on large scales}.  In terms of our constraint [1.7],
this result implies that we should replace the constant parameter $\delta$
with some function $\delta (x)$ which is a {\it decreasing} function
of time (and hence a decreasing function of $x$) during the structure
producing portion of inflation.  Keep in mind that in this present
discussion $\delta$ represents the upper bound on the density
fluctuations and not the amplitude of the fluctuations themselves.

For this discussion, we take the spectrum of density fluctuations
to be a simple power-law (see equations [3.8] and [3.9]).
We can incorporate this scale dependence into our density
perturbation constraint by writing it in the form
$${1 \over 10} {H^2 \over {\dot \phi} } \le
\delta (x) = \delta_0 \exp [ - \alpha x ] \, , \eqno(5.4)$$
where $\delta_0$ represents the size of the allowed
perturbations at the largest size scale (the present-day
horizon scale) and where we have defined
$$\alpha = N \, (1 - n) /  2 \, . \eqno(5.5)$$
As before, the scale invariant spectrum $n = 1$ corresponds
to $\alpha = 0$ while a spectrum with more power on
large scales $n < 1$ corresponds to $\alpha > 0$.
Notice that we have written equation [5.4] as an
inequality; we assume that the perturbations produced
during inflation (left-hand-side of the equation) are
smaller than (or equal to) the actual primordial perturbations
(right-hand-side of the equation).
In terms of the dimensionless functions introduced
in Sec. II, the density perturbation constraint
becomes
$${p(x) \over h(x) } \, \ge \, \exp[\alpha x] \, .
\eqno(5.6)$$
In previous work [2], we assumed that $\alpha = 0$
(i.e., that $\delta$ is constant).  For positive values
of $\alpha$ (i.e., for $n < 1$), our new constraint is
more restrictive than that used previously and hence
leads to tighter constraints.

We now show how this more restrictive constraint
affects our upper bound on the fine-tuning parameter
$\lft$.   To obtain a bound on $\lft$, we must find
an upper limit to the functional $\jfun$ defined in
equation [2.15].  This functional is maximized by
choosing $h$ so as to saturate the density perturbation
constraint, which now takes the form of equation [5.6];
we thus obtain
$$\jfun [p, h] \le {\int_0^1 p^4 \, dx \over
\Bigl[ \int_0^1 p \exp[ \alpha x ] \, dx \Bigr]^4 }
\ \ \equiv \,  \jfun_\alpha [p] , \eqno(5.7)$$
where the function $p(x)$ appearing in the functional
$\jfun_\alpha$ is subject to the same constraints as before.
For the case of $\alpha > 0$ ($n < 1$), the denominator
in equation [5.7] is clearly larger than for $\alpha = 0$
($n=1$); thus it is immediately clear
that $\jfun_\alpha < \jfun_0$ for any
nonzero value of $\alpha$, where $\jfun_0$ is the ``old''
functional with $\alpha = 0$. Furthermore, our intuition tells
us that $\jfun_\alpha$ reaches its maximum value when the
function $p(x)$ is as nonuniform as possible; in other words,
we suspect that the least upper bound for $\jfun_\alpha [p]$
occurs for a function $p$ that saturates the overdamping
constraint.  We show in Appendix B that this conjecture
is in fact correct.  The maximum for $\jfun_\alpha$ occurs for
the function $p(x)$ which {\it decreases} as fast as
possible given the overdamping constraint.  We thus obtain
the bound
$$\jfun_\alpha \le {81 N^3 \over 32} \bigl[ 1 -
2 \alpha/3N \bigr]^4 \, {1 - {\rm e}^{-6N} \over
\bigl[ 1 - {\rm e}^{-3N/2 + \alpha} \bigr]^4 } \approx
{81 N^3 \over 32} \bigl[ 1 - 2 \alpha/3N \bigr]^4
\, , \eqno(5.8)$$
where the final approximate equality holds for most
cases of interest since the exponentials are small.
For $\alpha \ne 0$, this bound is tighter
than that obtained previously in Ref. [2];  however, the
factor $\cal F$ by which the bound is tighter is rather small,
$${\cal F} \approx \bigl[ 1 - 2 \alpha/3N \bigr]^{-4} \, = \,
\Bigl( {3 \over 2 + n } \Bigr)^4 \, , \eqno(5.9)$$
where we have used the definition of $\alpha$ in the second
expression.  Thus, for the largest expected departures from scale-invariance,
$n \sim 1/2$, we find $\cal F$ $\approx$ 2.  Even for the rather
extreme departure from scale-invariance of $n=0$, we obtain only a
modest increase in the bound with $\cal F$ = 81/16 $\approx 5$.  We
therefore conclude that departures from scale-invariance lead to
moderately tighter constraints on the fine-tuning parameter $\lft$.

\goodbreak \bigskip
\centerline{\bf VI. SUMMARY AND DISCUSSION}
\nobreak \medskip

In this paper, we have found constraints on the scalar field
potential for a general class of inflationary models which
have slowly rolling fields.  These constraints apply to
all models of inflation which exhibit overdamped motion of
the scalar field and which obey the density perturbation
constraint.  This work thus extends that of Ref. [2].

\item{[1]}
We have studied the reconstruction of the inflationary
potential by considering both scalar and tensor modes.
The simultaneous consideration of both types of perturbations
leads to a differential equation which could be solved to find
the potential if the total primordial spectrum of perturbations
were known (see also Ref. [15]).  We showed how constraints on this
spectrum imply corresponding constraints on the reconstructed
potential $V(\phi)$.  Figures 1 -- 4 show the reconstructed
potentials for the expected range of parameter space.
For the case of density perturbation spectra with moderate
departures from scale invariance (e.g., $n = 0.6)$ and
little contribution from tensor modes, the reconstructed
potential is very similar to a cosine (see Figure 5) such
as in the model of Natural Inflation [7].

\item{[2]}
We have derived a relationship between the amplitude of
tensor perturbations and the width of the scalar field
potential (see equation [3.21]).  In particular, the average
ratio $\rave$ of tensor to scalar perturbations is comparable
to the dimensionless width $\Delta \phi/\mp$ of the potential.
Thus, if tensor perturbations are important, then the width
of the potential must be comparable to the Planck mass.
As we discuss in item [3] below, the width $\Delta \phi$
is bounded from above; as a result, the average ratio
$\rave$ is also bounded from above. This result implies
that while scalar perturbations can dominate over
tensor perturbations by an arbitrarily large factor,
the converse is not true:  tensor perturbations can
be {\it at most} a factor of $\sim$1.6 larger than
scalar perturbations.

\item{[3]}
We have found both upper and lower limits on the change $\Delta \phi$
of the scalar field during the phase of inflation which produces
cosmic structure (see equation [4.14]).  These limits can be
summarized by the relation
$$0.4 \, \Bigl( {\Lambda \over 10^{17} {\rm GeV} } \Bigr)^2
\, \le \, {\Delta \phi \over \mp} \, \le \, 3.9 \,
\Bigl( {N \over 8} \Bigr)  \, . $$
The lower limit depends on the energy scale at which inflation
takes place.  For energy scales larger than the GUT scale, the
width $\Delta \phi$ must be larger than the Planck scale $\mp$.
The upper limit implies that the change in the scalar field
during the $N=8$ e-foldings of structure-forming perturbations
cannot be larger than $\sim$4 $\mp$.

\item{[4]}
We have found both upper and lower bounds on the change
$\Delta V$ of the potential during inflation.  These
bounds can be used to find an upper limit on the energy
scale $\Lambda$ of the part of inflation when cosmological
structure is produced (see equation [4.31]),
$${\Lambda \over \mp } \le 6 \times 10^{-3} \,
\Bigl( {\delta \over 2 \times 10^{-5} } \Bigr)^{1/2} \, , $$
where $\delta$ is the maximum allowed amplitude of density
perturbations (see equation [1.4]).  This limit shows that
the epoch of structure-forming perturbations must take place
at an energy scale less than about the GUT scale.  This bound
is almost identical to those found earlier from the consideration
of scalar perturbations [16, 18]. The bound is comparable to
that obtained from the consideration of tensor perturbations [19].

\item{[5]}
We have also presented a very rough argument which indicates
that inflation at very low energy scales will encounter some
difficulty: the fractional change in the height of the potential
during the $N = 8$ e-foldings of structure formation is very
small when the energy scale $\Lambda$ is small, i.e.,
$$\eta = {\Delta V \over V_B} \sim 6 \times 10^{29}
\Bigl( {\Lambda \over \mp} \Bigr)^4 \sim {1 \over 10}
\Bigl( {\Lambda \over 10^{12} {\rm GeV} } \Bigr)^4 \, . $$
If $\Lambda$ is small compared to $\sim 10^{12}$ GeV,
then $\eta \ll 1$ and it is difficult for the potential
to drop to (roughly) zero in the remaining e-foldings
for a normally shaped potential.

\item{[6]}
We have found a {\it lower} bound on the fine-tuning
parameter $\lft$.  Our previous bound [2] showed
that the parameter $\lft$ must be quite small
$(\le 10^{-7})$; this new bound shows that $\lft$
cannot be made arbitrarily small.  These bounds thus
confine the fine-tuning parameter to the range
$$6 \times 10^{-11} \, \Bigl( {\Lambda \over 10^{17} {\rm GeV} }
\Bigr)^8 \le \, \lft \, \le \, 10^{-7} \, .  $$

\item{[7]}
We have explored the effects of non-scale-invariance
of density perturbations on the fine-tuning parameter
$\lft$ of Ref. [2].   If the density perturbations
are required to be non-scale-invariant, then we obtain
a stronger bound on $\lft$.  However, for the
departures from scale-invariance proposed as an explanation
of recent observations of cosmological data
on large scales (e.g., $n \approx 0.6$; see, e.g.,
Ref. [7]), the bound is improved by a rather modest
factor (${\cal F} \sim 2$).

\bigskip
\bigskip
\newpage
\centerline{\bf Acknowledgements}
\medskip

We would like to thank Ira Rothstein and Richard Watkins for
useful discussions and Douglas Adams for his invaluable assistance.
F. C. A. was supported by the NSF Young Investigator Program and
by NASA Grant No. NAGW-2802.  K. F. was supported by the
Presidential Young Investigator Program.

\vskip 1.0truein
\centerline{\bf APPENDIX A: GLOBAL BOUND ON $\Delta \phi$}
\nobreak \bigskip

In this appendix, we show that the lower bound of equation
[4.10] obtained for the functional $\kfun [p, h]$ is
in fact the greatest lower bound.  So far we have
shown that
$$\kfun [ p, h] \ge \kfun_\ast \, \eqno({\rm A}1)$$
where $\kfun_\ast$ is given by $\kfun$ evaluated
using $h (x) = p(x)$, which saturates the density
perturbation constraint, and
$$p(x) = p_\ast (x) \equiv \exp [ - 3Nx / 2 ] \, ,
\eqno({\rm A}2)$$
which saturates the overdamping constraint.

Suppose that $\kfun_\ast$ is not the greatest lower
bound.  Then there exist functions $p(x)$ and $h(x)$
which satisfy the constraints (and the initial conditions)
and for which $\kfun [p, h] < \kfun_\ast$. In other words,
$$\int_0^1 (p^2 / h) \, dx \, < \,
\int_0^1 \exp[-3Nx/2] \, dx \, . \eqno({\rm A}3)$$
Since both integrands are positive definite, it
follows that for some point $x_0$, we must have
$$ {p^2 \over h} \Big|_{x_0} \, < \,
\exp[-3N x_0 / 2 ] \, . \eqno({\rm A}4)$$
Since the functions $p$ and $h$ satisfy the
density perturbation constraint (by hypothesis),
$p/h \ge 1 $ for all $x$ and in particular for $x_0$.
Equation [A4] implies that
$$p(x_0) < \exp[-3N x_0 / 2 ] \, . \eqno({\rm A}5)$$
This final inequality violates the overdamping
constraint and thus leads to a contradiction.
Hence, the bound of equation [4.10] is the greatest
lower bound.

\newpage
\centerline{\bf APPENDIX B: BOUND ON THE FUNCTIONAL $\jfun_\alpha$}
\nobreak \vskip 0.10truein

In this Appendix, we find the least upper bound on
the functional $\jfun_\alpha [p]$ defined by equation
[5.7] in the text.  In particular, we
show that the desired bound is given by the functional
evaluated at $p=p_\ast(x)$, where $p_\ast (x)$ is the
function which decreases as fast as possible while
maintaining the overdamping constraint [2.17], i.e.,
$$p_\ast (x) = \exp [-3N x / 2] \, , \eqno({\rm B}1)$$
where we have used the initial condition [2.19].

We present a proof by contradiction.  Suppose that
$p_\ast$ does not provide the least upper bound as
claimed.  Then there exists another function $p(x)$
which satisfies the constraints of Sec. II and for
which
$$\jfun_\alpha [p] \ge \jfun_\alpha [p_\ast] \, .
\eqno({\rm B}2)$$
After introducing the notation
$$\langle \dots \rangle \equiv \int_0^1 (\dots) \, dx
\, , \eqno({\rm B}3)$$
we can write the condition [B2] in the form
$${ \langle p^4 \rangle \over
\langle p \exp[\alpha x] \rangle^4 } \ge
{ \langle p_\ast^4 \rangle \over
\langle p_\ast \exp[\alpha x] \rangle^4 } \, .
\eqno({\rm B}4)$$
However, we also know that $p_\ast$ does provide
the least upper bound for the functional $\jfun$
(as shown in Ref. [2]).  As a result, the functions
$p(x)$ and $p_\ast (x)$ must also obey the inequality
$${ \langle p^4 \rangle \over \langle p \rangle^4 } \le
{ \langle p_\ast^4 \rangle \over \langle p_\ast \rangle^4 }
\, . \eqno({\rm B}5)$$
The combination of these two results can be put in
the form
$${ \langle p \exp[\alpha x] \rangle^4 \over
\langle p_\ast \exp[\alpha x] \rangle^4 } \, \le \,
{\langle p^4 \rangle \over \langle p_\ast^4 \rangle }
\, \le \, { \langle p \rangle^4 \over \langle p_\ast \rangle^4 }
\, . \eqno({\rm B}6)$$
Eliminating the middle portion of equation [B6] and
taking the fourth root, we find
$${ \langle p \exp[\alpha x] \rangle \over
\langle p_\ast \exp[\alpha x] \rangle } \, \le \,
{ \langle p \rangle \over \langle p_\ast \rangle }
\, . \eqno({\rm B}7)$$
Next we define ``weight functions'' according to
$$w(x) \equiv {p(x) \over \langle p \rangle} \, ,
\qquad {\rm and} \qquad w_\ast (x) \equiv
{p_\ast (x) \over \langle p_\ast \rangle} \, ,
\eqno({\rm B}8)$$
where, by definition,
$$\langle w \rangle = 1 = \langle w_\ast \rangle
\, . \eqno({\rm B}9)$$
Using the weight functions, the inequality [B7]
becomes simply
$$\langle w \exp[\alpha x] \rangle \le
\langle w_\ast \exp[\alpha x] \rangle \, . \eqno({\rm B}10)$$
In what follows, we show that this equation is false and
hence leads to a contradiction.

Since both $w$ and $w_\ast$ have mean values of unity
(Eq. [B9]) and since the two functions are different,
we must have $w > w_\ast$ for some values of $x$ and
$w < w_\ast$ for other values of $x$.  However, by
construction, the function $w_\ast$ is monotonically decreasing
and is decreasing as fast as possible given the overdamping
constraint.  Therefore, $w = w_\ast$ at only one point [31],
denoted here as $x_0$ and
$$w(x) < w_\ast (x) \qquad {\rm for} \qquad
x < x_0 \, , \eqno({\rm B}11{\rm a})$$
$$w(x) > w_\ast (x) \qquad {\rm for} \qquad
x > x_0 \, . \eqno({\rm B}11{\rm b})$$
Thus,
$$\int_0^{x_0} (w_\ast - w) dx = C =
\int_{x_0}^1 (w - w_\ast) dx \, . \eqno({\rm B}12)$$
Now, using the Mean Value Theorem, we find
$$\int_0^{x_0} (w_\ast - w) \exp[\alpha x] \, dx =
\exp[\alpha \xi] \int_0^{x_0} (w_\ast - w)  \, dx =
\exp[\alpha \xi] \, C \, , \eqno({\rm B}13)$$
where $\xi$ is some number in the range $0 < \xi < x_0$.
Similarly, using the Mean Value Theorem a second time,
we find
$$\int_{x_0}^1 (w - w_\ast) \exp[\alpha x] \, dx =
\exp[\alpha \eta] \int_{x_0}^1 (w - w_\ast)  \, dx =
\exp[\alpha \eta] \, C \, , \eqno({\rm B}14)$$
where $\eta$ lies in the range $x_0 < \eta < 1$.
Notice that for $\alpha \ne 0$, neither $\xi = x_0$
nor $\eta = x_0$ so that $\eta > \xi$.  Putting these
results together, we find
$$\int_0^{x_0} (w_\ast - w) \exp[\alpha x] \, dx =
\exp[\alpha \xi] \, C \, < \exp[\alpha \eta] \, C \, =
\int_{x_0}^1 (w - w_\ast) \exp[\alpha x] \, dx \, .
\eqno({\rm B}15)$$
After a bit of rearrangement, this expression can be
written in the form
$$\langle w_\ast \exp[\alpha x] \rangle \, < \,
\langle w \exp[\alpha x] \rangle \, , \eqno({\rm B}16)$$
which contradicts equation [B10] above.  Thus, our
supposition that $p_\ast$ does not provide the
least upper bound leads to a contradiction.
The supposition must be false, i.e., the least
upper bound of the functional $\jfun_\ast [p]$
is given by the functional evaluated at
$p = p_\ast$.

\newpage
\vskip 1.0truein
\centerline{\bf REFERENCES}
\nobreak \vskip 0.10truein

\item{[1]}{A. H. Guth, {\it Phys. Rev.} D {\bf 23}, 347 (1981).}

\item{[2]}{F. C. Adams, K. Freese, and A. H. Guth,
{\it Phys. Rev.} D {\bf 43}, 965 (1991). }

\item{[3]}{A. H. Guth and E. Weinberg, {\it Nucl. Phys.}
{\bf B212}, 321 (1983); S. W. Hawking, I. G. Moss, and J. M.
Stewart, {\it Phys. Rev.} D {\bf 26}, 2681 (1982).}

\item{[4]}{A. D. Linde, {\it Phys. Lett.} {\bf 108B}, 389 (1982).}

\item{[5]}{A. Albrecht and P. J. Steinhardt, {\it Phys. Rev. Lett.}
{\bf 48}, 1220 (1982).}

\item{[6]}{A. D. Linde, {\it Phys. Lett.} {\bf 129B}, 177 (1983).}

\item{[7]}{K. Freese, J. A. Frieman, and A. V. Olinto,
{\it Phys. Rev. Lett.} {\bf 65}, 3233 (1990);
F. C. Adams, J. R. Bond, K. Freese, J. A. Frieman,
and A. V. Olinto, {\it Phys. Rev.} D {\bf 47}, 426 (1993).}

\item{[8]}{E. W. Kolb and M. S. Turner,
{\it The Early Universe} (Frontiers in Physics;
Vol. 69, Addison Wesley, 1990).}

\item{[9]}{D. La and P. J. Steinhardt, {\it Phys. Rev. Lett.}
{\bf 376}, 62 (1989); D. La and P. J. Steinhardt,
{\it Phys. Lett.} {\bf 220B}, 375 (1989).}

\item{[10]}{F. C. Adams and K. Freese, {\it Phys. Rev.} D
{\bf 43} 353, (1991).}

\item{[11]}{A. A. Starobinskii, {\it Phys. Lett.} {\bf 117B}, 175
(1982); A.~H.~Guth and S.-Y.~Pi, {\it Phys. Rev. Lett.} {\bf 49},
1110 (1982); S.~W.~Hawking, {\it Phys. Lett.} {\bf 115B}, 295
(1982); J.~M.~Bardeen, P.~J.~Steinhardt, and M.~S.~Turner, {\it
Phys. Rev.} D {\bf 28}, 679 (1983); and R. Brandenberger, R.
Kahn, and W. H. Press, {\it Phys. Rev.} D {\bf 28}, 1809 (1983).}

\item{[12]}{L. Abbott and M. Wise, {\it Nucl. Phys.}
{\bf B 244}, 541 (1984);
A. A. Starobinskii, {\it JETP Lett.} {\bf 11}, 133 (1985);
M. White, {\it Phys. Rev.} D {\bf 46}, 4198 (1992).}

\item{[13]}{G. Smoot et al., {\it Astrophys. J. Lett.}
{\bf 396}, L1 (1992); E. L. Wright et al.,
{\it Astrophys. J. Lett.} {\bf 396}, L3 (1992).}

\item{[14]}{P. J. Steinhardt and M. S. Turner,
{\it Phys. Rev.} D {\bf 29}, 2162 (1984).}

\item{[15]} H. M. Hodges and G. R. Blumenthal, {\it Phys. Rev.}
D {\bf 42}, 3329 (1990);
D. S. Salopek, {\it Phys. Rev. Lett.} {\bf 69}, 3602 (1992);
F. Lucchin, S. Matarrese, and S. Mollerach,
{\it Astrophys. J. Lett.} {\bf 401}, 49 (1992);
E. J. Copeland, E. W. Kolb, A. R. Liddle, and J. E. Lidsey,
{\it Phys. Rev. Lett.} {\bf 71}, 219 (1993);
E. J. Copeland, E. W. Kolb, A. R. Liddle, and J. E. Lidsey,
{\it Phys. Rev.} {D}, {\bf 48}, 2529 (1993);
A. R. Liddle and D. H. Lyth, {\it Phys. Lett.} {\bf 291B}, 391 (1992);
R. L. Davis, H. M. Hodges, G. F. Smoot, P. J. Steinhardt, and M. S. Turner,
{\it Phys. Rev. Lett.} {\bf 69}, 1856 (1992);
M. S. Turner, {\it Phys. Rev.} {\bf D}, in press (1993).

\item{[16]}{D. H. Lyth, {\it Phys. Lett.} {\bf 147 B},
403 (1984).}

\item{[17]}{D. H. Lyth, {\it Phys. Lett.} {\bf 246 B},
359 (1990).}

\item{[18]}{S. W. Hawking, {\it Phys. Lett.} {\bf 150 B},
339, (1985).}

\item{[19]}{L. M. Krauss and M. White, {\it Phys. Rev. Lett.}
{\bf 69}, 869 (1992).}

\item{[20]} E. J. Copeland, E. W. Kolb, A. R. Liddle, and J. E. Lidsey,
{\it Phys. Rev.} {\bf D}, in press (1993).

\item{[21]} A similar differential equation is derived but not solved
analytically in the paper by Copeland et al. (1993) in Ref. [15].

\item{[22]} For further discussion of density perturbations,
see standard textbooks such as Refs. [23] and [8]; see also
the recent review of Ref. [24].

\item{[23]} P.J.E. Peebles, {\it Principles of Physical
Cosmology} (Princeton University Press, 1993).

\item{[24]} A. R. Liddle and D. H. Lyth, {\it Physics Reports},
{\bf 231}, 1 (1993).

\item{[25]} We note that different authors use different
normalizations for $\delta_S$ and $\gwper$ (see Refs. [15]), i.e.,
they use different numerical coefficients for the expressions in
equations [3.1] and [3.2].  If we use these different normalizations,
the constant $C$ will change by a factor of order unity.  However,
this change is sufficiently small that it will not affect the
present discussion; in particular, the resulting shapes of the
potential will not change.

\item{[26]} R. Crittenden, J. R. Bond, R. L. Davis,
G. Efstatthiou, and P. J. Steinhardt, {\it Phys. Rev. Lett.}
{\bf 71}, 324 (1993).

\item{[27]} If we had the opposite case, $\phi_1 < \phi_2$,
then we could simply perform a reflection $\phi \to - \phi$
without changing the physics.

\item{[28]} Strictly speaking, the energy density can become
slightly negative;  present day cosmological constraints allow
a negative vacuum energy density roughly comparable to the
critical density at the present epoch ($\sim$ $8 h^2 \times$
$10^{-47}$ GeV$^4$).  This correction will not affect the
present argument in any substantial way.

\item{[29]} We note that the $\Delta V$ in the numerator
refers to the change in the scalar field potential, while
the $V$ in the denominator refers to the {\it total}
vacuum energy density.  For inflationary models with
only one scalar field, the two potentials are the same.

\item{[30]} See, e.g., R. Y. Cen, N. Y. Gnedin, L. A. Kofman,
and J. P. Ostriker, {\it Astrophys. J. Lett.} {\bf 399}, L11
(1992); R. Y. Cen and J. P. Ostriker, {\it Astrophys. J.}
{\bf 414}, 407 (1993); see also Ref. [7].

\item{[31]} It is possible for $w = w_\ast$ to hold for a finite
interval rather than for a single point.  However, the same argument
can be easily modified to accommodate this possibility.
%

\newpage
\vskip1.0truein
\centerline{\bf FIGURE CAPTIONS}
\bigskip

\medskip
\noindent
Figure 1. Reconstructed inflationary potential for the
case where tensor perturbations provide 31\% of the
total at $x=0$ (i.e., $\vr_B = 10^{-1}$).  [Note that
the parameter $x$ characterizes the number of e-foldings
subsequent to the epoch $x=0$, which occurs $\sim$60
e-foldings before the end of inflation when cosmological
structure on the scale of our horizon was produced.]
The various curves are for indices $n$ = 0.5,
0.6, 0.7, 0.8, 0.9, and 1.0 (from bottom to top).
The open symbols denote the epoch at which galaxy-sized
perturbations leave the horizon during inflation.

\medskip
\noindent
Figure 2. Reconstructed inflationary potential for the
case where tensor perturbations provide 10\% of the
total at $x=0$ (i.e., $\vr_B = 10^{-2}$).
The various curves are for indices $n$ = 0.5,
0.6, 0.7, 0.8, 0.9, and 1.0 (from bottom to top).
The open symbols denote the epoch at which galaxy-sized
perturbations leave the horizon during inflation.

\medskip
\noindent
Figure 3. Reconstructed inflationary potential for the
case where tensor perturbations provide 3.1\% of the
total at $x=0$ (i.e., $\vr_B = 10^{-3}$).
The various curves are for indices $n$ = 0.5,
0.6, 0.7, 0.8, 0.9, and 1.0 (from bottom to top).
The open symbols denote the epoch at which galaxy-sized
perturbations leave the horizon during inflation.

\medskip
\noindent
Figure 4. Reconstructed inflationary potential for the
case where tensor perturbations provide 1\% of the
total at $x=0$ (i.e., $\vr_B = 10^{-4}$).
The various curves are for indices $n$ = 0.5,
0.6, 0.7, 0.8, 0.9, and 1.0 (from bottom to top).
The open symbols denote the epoch at which galaxy-sized
perturbations leave the horizon during inflation.

\medskip
\noindent
Figure 5. Comparison of reconstructed inflationary
potential and a cosine potential.  The reconstructed
potential was obtained using $n = 0.6$ and
$\vr_B = 10^{-4}$ (tensor perturbations initially
produce 1\% of the total).  The fit was obtained by
constraining the cosine curve to agree with the
reconstructed potential at the endpoints $x=0$
and $x=1$.

\bye